\documentstyle[preprint,prl,aps]{revtex}
\begin{document}
\preprint{\vbox{\hbox{July 1994}\hbox{IFP-708-UNC}\hbox{VAND-TH-94-15}}}
\draft
\title{Massive Superstrings Are Black Holes}
\author{\bf Paul H. Frampton$^{(a)}$ and Thomas W. Kephart$^{(a,b)}$}
\address{(a) Institute of Field Physics, Department of Physics
and Astronomy,\\
University of North Carolina, Chapel Hill NC  27599-3255}
\address{(b)   Department of Physics
and Astronomy,\\
Vanderbilt University, Nashville, TN  37235\footnote{Permanent address}}
\maketitle

\begin{abstract}
We argue that it is inconsistent to ignore the gravitational backreaction
for on-shell superstring states at the Planck mass and beyond, and that
these quantum states become Kerr black holes in the classical limit.
Consequences are discussed.
\end{abstract}

\newpage
     In assessing progress towards a satisfactory theory of
 quantum gravity it is important to
 evaluate the present status of, and relationship between,
the principal disparate
yet overlapping approaches: (i) the operator, canonical and path
integral approaches \cite{Hawk}
are the most direct and historically first attempts at quantum gravity based
on techniques used so successfully for quantum field theory in flat spacetime;
(ii) the dilemmas and paradoxes surrounding black hole evaporation and
thermodynamics, including backreaction, provide a fruitful direction toward
quantum gravity \cite{Hawk2}; (iii) superstring theory in some version yet
to be identified promises a basis for quantum gravity  and its connection
to particle theory {\cite{Fram},\cite{GSW}};
and (iv) the ideas within quantum cosmology
about a satisfactory boundary condition to characterize the big bang provide
an equally valid mode of investigation[5].

Here we shall concentrate on the relationship between (ii) and (iii),
black holes and superstrings and their thermodynamics at the
Planck scale.  Toward the end of the Letter we  shall comment about their
impact on (iv), the preinflationary big bang.

The string spectrum as determined by factorization of the tree amplitudes
comprises, to lowest order, an infinite set of parallel linear
Regge trajectories (Fig. 1).  The leading quantum trajectory
for the superstring is \cite{GSW}
\begin{equation}
{\alpha_{graviton}}(s) = 2 + {\alpha}'s
\end{equation}
with other trajectories at
\begin{equation}
{\alpha }(s) = 2 + {\alpha}'s - n/2
\end{equation}
for $n = 1,2,...$.  The massless spectrum comprises the graviton, the
gravitino,  and lower spin states, e.g., gauge bosons, quarks and leptons.

Massive states have masses at least comparable to the Planck
mass $\sim 10^{-19} GeV$.  For the leading
trajectory (1), $J \sim {\alpha}' M^2$ as $M^2 \rightarrow \infty$.
What is the slope ${\alpha}'$?  For the leading
trajectory a classical model is of a straight string of length
$2l$ and density $\rho$ per unit length.  Using the boundary condition
that the ends move at the speed of light $(c)$ the value of $J$
is
\begin{equation}
\int_{-l}^{l} dx {\rho}(xc/l) x = {cM^2}/{6{\rho}}
\end{equation}
and $\rho $ is the ratio $(Planck Mass)/(Planck Length) = ({\hbar}c/G)^{1/2}
({\hbar}G/c^3)^{-1/2} = {c^2}/G$.  Thus the superstring slope is roughly
\begin{equation}
{{\alpha}'}_{superstring} = G/{6c}
\end{equation}
In scattering processes it is possible to show that the states with
$J \leq (M^2 {\alpha}')^{1/2}$ are significantly more strongly
coupled \cite{FN} although all states with $J \leq  M^2 {\alpha}'$ do
occur.

It is interesting to compare
the string slope, eq (4) to the condition relating $J$ and
$M$ for a Kerr black hole \cite{K}.  The Kerr solution without
charge has an event horizon at radius
\begin{equation}
R(Kerr) = GM/c^2 + \sqrt{G^2 M^2/c^4 - J^2/M^2 c^2}
\end{equation}
in $d=4$ spacetime dimensions for angular momentum
${\hbar} J$.  Only states with
\begin{equation}
J \leq G M^2/c
\end{equation}
may form a black hole.  Thus all on-shell superstring
states below the line
\begin{equation}
\alpha = G M^2/c
\end{equation}
can and will collapse to black holes.  This is certainly true in the
classical limit where the black hole mass satisfies $M_{Planck} \ll \mu$; it is
not
clear where or if
the argument breaks down for  $\mu \rightarrow M_{Planck}$.
Comparing Eqs. (1) and (7) shows that one
expects {\it all} massive superstring states on mass shell to form black holes.
The Chew-Frautschi plot is depicted in Fig. 1.  In Eqs. (1), (2), and in
Fig. 1 backreaction can shift the black hole masses giving some non-linearity
to the Regge tragectories; this merits further study using the results
of \cite{HKW}.

	The superstring massive states are physical black holes only on mass
shell;
off-shell, as virtual states contributing to low-energy scattering of
massless particles, this does not detract from the usefulness of perturbative
superstrings as an effective theory.

	Accepting that massive superstring states describe black holes, we may
use a
toy dual model for scattering of two black holes.  It was shown twenty-five
years ago \cite{FN} that in a dual model for scattering two states the biggest
contribution comes from those states with angular momentum $J \leq (constant)
M$
as one would expect from a lever-arm argument for a given impact parameter.
There, one started with a hadronic amplitude
\begin{equation}
A(s,t) = {\Gamma}(1 - {\alpha} (s)) {\Gamma}(1 - {\alpha} (t))/
{\Gamma}(1 - {\alpha} (s) - {\alpha} (t))
\end{equation}
and computed the asymptotic form of the partial wave amplitudes; we shall
go into more details below, where we shall assume Eq. (8) is applicable to
spinless black-hole scattering.

	A connection between superstrings states and black holes can
be made in
a different fashion.  The Helmholtz free energy per unit 4-volume F/VT in a
thermal distribution of strings at temperature T can be evaluated from world-
sheet path integrals for string propagation at the one-loop level \cite{P}
for T
below the value $T_{critical}$ (not very different, but somewhat below,
$T_{Hagedorn}$
and $T_{Planck}$) where a phase transition may occur \cite{AW}. Above
$T_{critical}$
a
zero-loop contribution dominates F such that even the concept of temperature
necessarily becomes ill-defined because gravitational instability disallows
consideration of a suitable large volume limit.

	The free energy from loop contributions and also from 1/N estimates
give,
at fixed effective coupling $g_{eff}^2$, the dependence \cite{AW}
\begin{equation}
F/VT \sim T^p
\end{equation}
where $p = (d-1)$ for field theory in d spacetime dimensions, and $p = 0$
 and $1$
for the cases of open and closed strings respectively.  This is how
it was suprisingly concluded that strings above the Hagedorn temperature have
far fewer degrees of freedom than a corresponding field theory.
To compare this superstring free energy to that calculated for black
holes,
we cite the generalized Helmholtz free energy \cite{Y} for the spinless
(Schwarzschild) limit.  This result rewritten for a dilute gas of black holes
is that
\begin{equation}
F = -4{\pi} M^2 T
\end{equation}
which can be rewritten as
\begin{equation}
F/VT = - 1/16 {\pi} T_H^2 V
\end{equation}
so that for fixed-size black holes the free energy per unit 4-volume
corresponds (in first approximation) to a constant value independent of
the temperature of the surroundings; that is, $p=0$
in Eq. (8).  This behavior is similar to that of the open string.  It is
interesting to note that for closed strings \cite{AW} where $p=1$ there is a
mysterious ultra-violet pathology in the classical limit $\hbar
\rightarrow 0$; it will
be interesting to investigate whether this smooth limit obtaining for open
strings carries over to their Kerr counterparts.

	Alternatively, by setting $T_H = T$ in Eq. (10), we find
\begin{equation}
F/VT \sim T^{-2}
\end{equation}
with an ${\hbar}^{-2}$ ultra-violet divergence in the classical limit.
For comparison
purposes it would be interesting to have a calculation of free energy for
a thermal distribution of membranes.

	If one accepts our thesis that massive superstrings on-shell are
Kerr
black holes, they must surely be quantum black holes.  This implies that
the black hole entropy should be computed along the general lines presaged
in {\cite{ZT},\cite{M}}.
Our result that massive superstrings are black holes
may be interestingly compared to work of 't Hooft \cite{tH} who gave
the complementary argument that black holes are strings.

	The spectrum of quantum black holes is predicted by the superstring to
be
${\sqrt{n}} M_{Planck}$
where $n$ is an integer.  This immediately
provides
predictions for the spectrum of emitted massless particles, for the entropy,
and for the evaporation.

	Suppose a black hole of mass ${\sqrt{n}} M_{Planck}$
 radiates a photon in decaying to a smaller
black hole of mass ${\sqrt{n'}} M_{Planck}$.
(A similar analysis could be carried out for decay
to other massless states \cite{Page}.)  Then the energy of the photon in the
rest
frame of the initial black hole is easily shown to be
\begin{equation}
E_{\gamma} = (M_{Planck}/2) ({n - n'})/{\sqrt{n}}
\end{equation}
This implies a characteristic spectrum of radiation, for example
\begin{equation}
{2 E_{\gamma}}/ {M_{Planck}} = k_1/{\sqrt{n}}, k_2/{\sqrt{n - k_1}}, ...,
k_n/{\sqrt{n - \sum_{i = 1}^{n - 1} k_i}}
\end{equation}
where the $k_i$ are integers.  For
$ n \rightarrow \infty$ this approximates a continuous spectrum,
but for small n the quantum nature becomes apparent and each photon has
of order the Planck energy in the rest frame of the decaying black
hole\cite{cosmic}.
The entropy is defined as the logarithm of the number of different quantum
configuration of the black hole.  As can be seen from Eq. (14), this
corresponds to the number of ordered partitions of $n$, that is $2^{n - 1}$
, and
consequently an entropy
\begin{equation}
S = (n - 1) {\ln{2}}
\end{equation}
This agrees with \cite{M}.

	To summarize, all mass-shell elementary particles are either massless
or
black holes.  The massless states are the superstring zero modes and include
all the familiar particles of the standard model, and possibly further
particles massless with respect to the Planck scale.  The black holes are
quantum states with spectrum $M(n) = {\sqrt{n}} M_{Planck}$
(up to backreaction corrections),
and are gravitationally confined superstrings.

	Let us pursue further the scenario to which we have been led.  During
the
very early universe these mini black holes can either evaporate into
massless states or, if the collsion time is sufficiently short compared
to the Hubble time and to the evaporation lifetime, they can also coalesce
into successively larger black holes.

	Returning to the toy dual model for black hole scattering, Eq. (8), we
may partial wave project the residue $F_n (t)$ at the level
${\alpha} (s) = n$ for Regge
intercept $a$ and external mass $\mu$ by writing

\begin{equation}
C_n^l = (l + 1/2) \int_{-1}^{+1} F_n (t(z)) P_l (z) dz
\end{equation}
and hence
\begin{equation}
C_n^l =    n (l + 1/2) (2 \pi)^{1/2} [(n - a - 4 {\mu}^2)/2]^{-1/2}J
\end{equation}
in which
\begin{equation}
J = 1/{2 \pi i} \int {d \zeta}/{\zeta}^{1/2}
     exp[(3a + 4 {\mu}^2 - 1) {\zeta}/2]I_{l + 1/2}[(n - a - 4 {\mu}^2 )
{\zeta}/2]/[2sinh(\zeta/2)]^{n+1}.
\end{equation}

Now by using asymptotic versions for the modified Bessel function for
$n,l \rightarrow \infty$ we
find that
\begin{equation}
C_n^l \sim {[{1/2}(l + 1/2)]^{1/2} / [({\zeta}_0 /2) \cosh{({\zeta}_0 /2)}}
- \sinh{({\zeta}_0 /2)}]^{1/2}E
\end{equation}
where
\begin{equation}
E = {[exp (\cosh{({\zeta}_0 /2 )} \tanh{({\zeta}_0 /4 )}]^{l + 1/2} }
/ [2 \sinh{({\zeta}_0 /2)}]^n 2 \cosh[(3a + 4 {\mu}^2 - 1) {\zeta}_0/2]
\end{equation}
for $n - l$ even. (For $n - l$ odd the final $\cosh$ becomes a $\sinh$.)
	From this we expect for black hole scattering a dominance by
intermediate Schwarzschild-like
black holes with $l \leq \sqrt{n}$, and of coalescence
into such black holes.  The
probability for such fusion can be estimated using the black hole lifetime
${\tau}_H \sim {M^3}/{n_{\phi}}$
and cross-section $\sigma \sim {M^2}/{\beta}$
where ${n_{\phi}}$ is the number of distinct massless modes
and $\beta$ is the black hole velocity.

	If the universe undergoes inflation after the big bang, then the
number of
primordial black holes remaining today will be diluted to a few per horizon
size provided that the early universe was not matter dominated by the
black holes to the extent that overclosure and recollapse occurred.  We
may check from a toy calculation that this dilution occurs, assuming
that all black holes started at the Planck time $t_1 = 3 \times {10}^{-44} s$
with $\mu = M_{Planck}$.  (It is not
clear that there cannot be a spectrum of black hole masses resulting from
string winding modes, or a thermal distribution there for $R < L_{Planck}$
\cite{BV}.)
At $t = t_1$ the horizon volume is estimated as $V_1 = {10}^{-9} {cm}^3$
and critical density is ${\rho}_1 = 2.5 \times {10}^{93} g/{{cm}^3}$.
Since the black hole lifetime varies as $M^3$
we take time steps $t_n = (8)^{n - 1} \times 3 \times {10}^{-44} s$
at which ${\rho}_n = (64)^{n - 1} \times {\rho}_1$
   This approach approximates the quantum
nature of mini black holes discussed
above.  The black hole mass is ${\mu}_n = (2)^{n - 1} \times {10}^{-5} g$
due to successive approximate doublings by fusion
with cross-section ${\sigma}_n = {(4)^{n - 1}{\sigma}_1}/ {\beta}_n^2 =
(4)^{n - 1} \times {10}^{-66} {cm}^3/{\beta}_n^2$  .
The probability $p_n$ of a fusion during the $n^{th}$
stage is
\begin{equation}
p_n = {{\sigma}_n {\tau}_n{\beta}_n{\rho}_n}/{\mu}_n
\end{equation}
where ${\beta}_n$ is the velocity ${\beta}_n^2 = (4 \sqrt{2})^{n - 1}$
and ${\rho}_n $ is the black hole
density ${\rho}_n$$ =$$ {\rho}_{n -1} ({128})^{-1} p_{n - 1}$.  Thus
\begin{equation}
p_n = 2^{-7/4} (p_{n-1})^2 = 2^{-a_n}
\end{equation}
where
\begin{equation}
a_n = 2 a_{n - 1} + 7/4 = 2^n (15/8) - 7/4
\end{equation}
We find that after seven 8-foldings corresponding to $T \sim {10}^{16} GeV$
 all the primordial
black holes have evaporated as anticipated.
	Thus primordial black holes do not survive past the GUT scale.  Black
holes can reform during cosmological phase transitions \cite{NP}, but it is
again ruled out that they survive evaporation to seed structure formation.

To conclude, superstrings are at best useful as a low-energy effective field
theory where the massive black holes are off-shell.  Any more complete
theory for higher (Planckian) emerges must involve massive superstrings
as black holes including their backreaction.

	We both thank the Aspen Center for Physics for hospitality while
preparing
this article. This work was supported in part by the US Department of
Energy under Grant Numbers DE-FG05-85ER-40219 and DE-FG05-85ER-40226.

\bigskip
{\bf Figure Caption:}

Fig 1.  Chew-Frautschi plot showing the first seven Regge trajectories.  The
line
marked by an asterisk (*) is that
    below which resonances are strongly coupled (see Ref. 6); the line marked
by (**)
is the black hole bound.


\begin{references}

\bibitem{Hawk}
  S. W. Hawking, The Path-Integral Approach to Quantum Gravity, in
				{\it General Relativity:  An Einstein Centenary
Survey}, Editors S. W. Hawking
    and W. Israel, Cambridge University Press (1979), Chapter 15.

\bibitem{Hawk2}
 S. W. Hawking, Comm. Math. Phys.  {\bf 43}, 199(1975).

\bibitem{Fram}
  P. H. Frampton, {\it Dual Resonance Models}, Benjamin (1974); reprinted in
			 an extended form as {\it Dual Resonance Models and
Superstrings}, World
    Scientific (1986).

\bibitem{GSW}
  M. B. Green, J. H. Schwarz and E. Witten, {\it Superstring Theory},
Cambridge
    University Press (1986).

\bibitem{HH}
 J. B. Hartle and S. W. Hawking, Phys. Rev. {\bf D28}, 296(1983).

\bibitem{FN}
  P. H. Frampton and Y. Nambu, Asymptotic Behavior of Partial Widths in
    the Veneziano Model of Scattering Amplitudes, in {\it Quanta}, Festscrift
    for G. Wentzel, Editors: P. G. O. Freund, C. J. Goebel and Y. Nambu,
    University of Chicago Press (1970), page 403.

\bibitem{K}
  R. Kerr, Phys. Rev. Lett. {\bf 11}, 327(1963);\\
 S. Chandrasekhar, {\it
The Mathematical
    Theory of Black Holes}, Oxford University Press (1983).

\bibitem{HKW}
J. W. York, Jr, Phys. Rev. {\bf D31}, 775(1985);\\
D. Hochberg and T. W. Kephart, Phys. Rev. {\bf D47}, 1465(1993);\\
D. Hochberg, T. W. Kephart and J. W. York, Phys. Rev. {\bf D48}, 479(1993);\\
D. Hochberg, T. W. Kephart and J. W. York, Phys. Rev. {\bf D49}, 5257(1993).

\bibitem{P}
     J. Polchinski, Comm. Math. Phys. {\bf 104}, 37(1986).

\bibitem{AW}
    J. J. Atick and E. Witten, Nucl. Phys. {\bf B310}, 291(1988).

\bibitem{Y}
     J. W. York, Phys. Rev. {\bf D33}, 2092(1986).


\bibitem{ZT}
     W. H. Zurek and K. S. Thorne, Phys. Rev. Lett {\bf 54}, 2171(1985).

\bibitem{M}
    V. F. Mukhanov, The Entropy of Black Holes, in {\it Complexity, Entropy
and
    the Physics of Information}, SFI Studies in the Science of Complexity,
    Vol. 8, Editor W. H. Zurek, Addison-Wesley (1990), page 47.

\bibitem{tH}
     G. 't Hooft, Nucl. Phys. {\bf B335}, 138(1990).

\bibitem{Page}
      D. N. Page, Phys. Rev. {\bf D13}, 198(1976).

\bibitem{cosmic}
A quantum spectrum for black holes/ strings of this type has very
interesting implications if some of these objects are present today.
Assuming they lose energy through two-body decays, they would produce extremely
energetic cosmic rays ( approx. $10^{28}$ eV) which could then degrade by
scattering from the cosmic background radiation to produce the high energy end
of the observed cosmic ray
spectrum (approx. $10^{20}$ eV), for which there there is at present no
conventional physics explanation.


\bibitem{BV}
     R. Brandenberger and C. Vafa, Nucl. Phys. {\bf B316}, 371(1989).

\bibitem{NP}
     P. N. Nasel'skii and A. C. Poharev, Sov. Astron. {\bf 29}, 487(1985).

\end{references}
\end{document}